\documentclass[aps,twocolumn,showpacs,floatfix,amssymb]{revtex4}
\usepackage{graphicx}
\begin{document}

\title{Optical and dc conductivities of  cuprates:\\
Spin-fluctuation scattering in the t-J model }

\author{ A. A. Vladimirov$^{a}$, D. Ihle$^{b}$, and N. M. Plakida$^{a,c}$ }
\affiliation{ $^a$Joint Institute for Nuclear Research, 141980 Dubna, Russia\\
$^{b}$ Institut f\"{u}r Theoretische Physik, Universit\"{a}t
Leipzig,
 D-04109, Leipzig, Germany \\
 $^{c}$Max-Planck-Institut f\"{u}r Physik komplexer Systeme, D-01187  Dresden,
Germany}

\date{\today}

\begin{abstract}
A microscopic theory of the electrical conductivity
$\sigma(\omega)$ within the $t$--$J$ model is developed. An exact
representation for $\sigma(\omega)$ is obtained using  the
memory-function technique for the relaxation function in terms of
the Hubbard operators, and the generalized Drude law  is derived.
The relaxation rate due to the decay  of charge excitations into
particle-hole pairs assisted   by antiferromagnetic spin
fluctuations is calculated in the mode-coupling approximation.
Using results for the spectral function of spin excitations
calculated previously,  the relaxation rate and the optical and
dc conductivities are calculated in a broad region of doping and
temperatures. The  reasonable agreement of the theory with
experimental data for cuprates proves the important role of
spin-fluctuation scattering in the charge dynamics.
\end{abstract}

\maketitle

\section{Introduction}

Studies of charge dynamics in superconducting cuprates provide
valuable information concerning  electron interaction with
bosonic modes which is important for elucidating the pairing
mechanism in high-temperature superconductors.  There is a vast
literature devoted to these studies (for  reviews see,
e.g.,~\cite{Timusk92,Basov05,Basov11,Plakida10}). Two major
scenarios have been proposed, the electron-phonon  coupling and
electron interaction with  the antiferromagnetic (AF) spin
fluctuations. Angle-resolved photoemission spectroscopy  (ARPES)
points to an important role of spin fluctuations in a
renormalization of the single-electron excitation spectrum (see,
e.g.,~\cite{Kordyuk10} and references therein) which is supported
by measurements of the infrared (IR) absorption in a wide region
of temperatures and doping  (see,
e.g.,~\cite{Hwang06,Yang09,Heumen09} and references therein). The
main argument against the spin-fluctuation pairing mechanism, a
weak intensity of spin fluctuations at the optimal doping seen in
inelastic magnetic neutron scattering
experiments~\cite{Bourges98}, was dismissed in recent resonant
inelastic x-ray scattering~\cite{LeTacon11}. In a large family of
cuprates  AF paramagnon excitations with dispersions and spectral
weights similar to those of magnons in undoped cuprates were
found. However, a decisive role of the electron-phonon
interaction (EPI) has been claimed in a number of theoretical
studies (for a review see~\cite{Maksimov10}).
\par
The  optical conductivity (OC) $\sigma(\omega)$ of cuprates
reveals a complicated evolution with doping and temperature. The
undoped parent  compounds are AF insulators, where the OC exhibits
a peak at the charge-transfer energy  $\omega \lesssim 1.8 $~eV.
Under doping, the insulator-to-metal transition  occurs when the
charge-transfer gap is filled up with states and the spectral
weight is transferred to the lower energy, the Drude peak at
$\omega \rightarrow 0$ with a  width $\omega \lesssim
600$~cm$^{-1}$  and a broad mid-infrared (MIR) band at higher
energies, $\omega \lesssim 5000$~cm$^{-1}$. With increasing hole
concentration, the MIR absorption shifts to lower energies and
merges with the Drude contribution (see, e.g.,
Refs.~\cite{Uchida91,Uchida96,Onose04,Lee05,Padilla05}). The
Drude peak significantly narrows with decreasing temperature and
is attributed to the relaxation of coherent quasiparticles, while
the origin of the temperature independent MIR contribution is
still under discussion. In Ref.~\cite{Lupi09} the
metal-to-insulator transition  (MIT) was studied by measuring the
OC for the
underdoped Bi-based and YBCO-based compounds  
for hole concentrations from $\delta = 0.12$ to $\delta = 0.015$.
With decreasing hole concentration, the Drude peak at low
temperatures transforms into a far-infrared (FIR) band at
energies $\omega \gtrsim 200$~cm$^{-1}$ which acquires a gap at
the MIT for hole doping $\delta \lesssim 0.07$. Note that the
onset of the metallic phase occurs at a doping much higher than
$\delta \simeq 0.02$  at which  the AF long range order (LRO)
vanishes ~\cite{Iye92}.
\par
It is generally   believed  that  superconducting cuprates are
doped Mott-Hubbard (charge-transfer~\cite{Zaanen85}) insulators,
where the insulating phase of the undoped parent compounds
appears due to a strong Coulomb repulsion, Hubbard  $U > U_{c2}$,
where the critical value  $U_{c2}$ for the MIT is larger than the
electronic bandwidth $W$ (see, e.g.,~\cite{Lee06}). In this case
the AF LRO in the undoped compounds originates from the strong AF
exchange interaction characteristic for Hubbard systems. However,
it is also possible to explain the insulating phase as caused by
the AF energy gap induced by the AF LRO where the Coulomb
interaction plays a secondary role. In recent publications, this
problem was discussed by analyzing the OC for typical
electron-doped Nd$_{2-x}$Ce$_x$CuO$_4$ (NCCO) and hole-doped
La$_{2-x}$Sr$_x$CuO$_4$ (LSCO) compounds. In
Ref.~\cite{Comanac08}, the OC was calculated for the Hubbard
model  in the paramagnetic and the AF phases. Using the dynamical
mean-field theory (DMFT)~\cite{Georges96}, the  optical spectral
weight (given by the restricted sum rule with the integration
over energy in Eq.~(\ref{g3}) up to $\Omega = 0.8$~eV) was
calculated. Comparing the doping dependence of the theoretical
and measured spectral weights for the NCCO, LSCO and other
cuprate compounds, it was found that, in the paramagnetic phase,
the fitted $U$ is smaller than the critical value $ U_{c2} \sim
1.5\, W $ for yielding the Mott-Hubbard insulating phase. At the
same time, the AF phase provides the insulating state for the
undoped system at the fitted $U$. So, it was concluded that
antiferromagnetism is essential  in  producing the insulating
state.
\par
A different conclusion concerning the hole-doped cuprates, was
obtained in Refs.~\cite{Weber08,Weber10}.  The OC was  calculated
for a realistic three-band $p$-$d$ model for NCCO and LSCO using
the local density approximation combined with the
DMFT~\cite{Kotliar06}. It was found that, whereas for the
electron-doped NCCO the AF interaction is necessary to yield the
insulating undoped state,  the hole-doped LSCO belongs to the
Mott-Hubbard system, where the insulating state is due to strong
electron correlations but not to the  AF interaction. However, in
the DMFT the short-range AF correlations are neglected. As shown
in Ref.~\cite{Plakida07}, by taking them into account, we find a
much narrower Hubbard band $\widetilde{W}$ which leads  to a lower
critical parameter $U_{c2}$ that may put the electron-doped
compounds also into the class of Mott-Hubbard insulators.
Therefore, for a more accurate estimation of $U_{c2}$ the cluster
DMFT~\cite{Haule07a} should be used. By this method the OC was
calculated in the  $t$-$J$ model in  Ref.~\cite{Haule07}.  The
results  reproduced quite well the delicate changes of the
spectral weight transfer at the transition to the superconducting
phase observed in experiments (see,
e.g.~\cite{Carbone06,Bontemps06} and references therein). This
proves that the $t$-$J$ model captures the essential physics of
the low-energy excitations in cuprates.
\par
In the limit of strong correlations, extensive  numerical studies
of the  OC within the Hubbard model and $t$-$J$ model for finite
systems have been carried out. Earlier results are reported in
Refs.~\cite{Stephan92,Dagotto94,Tohyama95,Eder95,Jaklic00}. Due
to a small size of the clusters, only one- or two-hole motion was
considered. In that case the Drude peak cannot be resolved, and
its intensity  versus doping was studied by calculations of the
kinetic energy (see Eq.~(\ref{r7})). Several peaks  found in the
OC in the MIR region may be related to  local spin excitations at
$\omega \simeq 2J$ and to string excitations at higher energies.
This observation was confirmed in analytical studies of the
charge correlations in a weakly doped $t$-$J$ model using the
cumulant expansion within the Zwanzig-Mori projection
technique~\cite{Vojta98}. The peaks found in the OC in the energy
region  $\omega \simeq 2J$ were assigned to excitations due to
internal degrees of freedom of the spin-bag quasiparticles.
\par
In several studies, an important role of the EPI  resulting in
polaronic effects  was stressed. However, contradicting
explanations were proposed for the two absorption bands observed
in the MIR region, one at a lower energy near the FIR region  and
another at higher energies. In Ref.~\cite{Cappelluti07}, the OC
of one hole in the Holstein $t$-$J$ model using the DMFT was
calculated. It was  shown that the IR absorption is characterized
by the coexistence of a magnon peak at low energy and a broad
polaronic band at higher energy. The two absorption bands  were
explained  in Ref.~\cite{Mishchenko08} by the coupling of a hole
to two kinds of bosonic excitations. The lower energy peak at
$\omega \simeq 1000$~cm$^{-1}$ was ascribed to the phonon
sideband, while the higher energy peak at $\omega \simeq
4600$~cm$^{-1}$ was considered as the magnon sideband of the
lower peak. As discussed in Ref.~\cite{Vidmar09},  the two-peak
structure  in the MIR region may be explained by the coupling of
a doped hole to magnetic excitations. The low-energy peak
represents the local magnetic excitation, attached to the hole,
while the higher-frequency peak corresponds to the MIR band that
originates from coupling to spin-wave excitations, broadened and
renormalized by phonon excitations. Thus, the studies of the
Holstein $t$-$J$ model suggest that  the complicated absorption
structure experimentally found in the MIR region is caused by
magnetic excitations which are coupled to phonons  via doped
holes.
\par
In analytical  studies, the OC is frequently evaluated within the
simple electron-hole (bubble) diagram approximation  for the
current-current correlation function proposed by
Allen~\cite{Allen71}. The finite-temperature version of the Allen
approximation was derived in Ref.~\cite{Shulga91}. This method was
used in studies of the optical IR data within electron-phonon
models (see~\cite{Maksimov10}  and references therein) and
spin-fermion models (see, e.g.,~\cite{Abanov03,Chubukov03} and
references therein).
\par
A  general approach based on the Mori memory-function
method~\cite{Mori65}  for the calculation of the current-current
relaxation function was proposed in Refs.~\cite{Goetze72}. In
this method, the transport relaxation rate is expressed directly
in terms of the force-force relaxation function which can be
further evaluated by  perturbation theory with a proper
consideration of the wave-vector dependence of the transport
vertex. In particular, in Ref.~\cite{Ihle94} the  memory-function
method was used to calculate the OC in the limit of strong
electron correlations within the Emery model for CuO$_2$ plane.
The relaxation rate for electrons scattered by AF spin
fluctuations was calculated in a fair agreement with experiments.
In Ref.~\cite{Plakida97}, by taking into account only the
incoherent part of the electronic spectrum, a scaling expression
for the frequency dependence of the relaxation rate and the
conductivity in the $t$--$J$ model was obtained.
\par
Electron interactions with bosonic modes can be  revealed  in the
low-energy part of the OC and dc conductivity. To shed more light
on the scattering mechanism in cuprates, in the present paper we
calculate the optical and dc conductivities within the $t$--$J$
model.  The main goal of this work is to demonstrate that AF
spin-fluctuation scattering is the essential mechanism of the
low-energy charge dynamics of underdoped and optimally doped
cuprates.
\par
Using the memory-function method, we derive an equation for the
relaxation rate which is determined by the kinematic interaction
for the Hubbard operators and depends only  on the parameters of
the $t$--$J$ model: the hopping parameters and  the AF exchange
coupling. The relaxation rate is calculated by taking into
account electron scattering by spin fluctuations  which are
described by  the spin-excitation spectral function calculated in
our previous works~\cite{Vladimirov09,Vladimirov11}. Therefore,
we are able to consider effects of spin excitations on the charge
dynamics within a microscopic theory without fitting parameters.
In our approach we  obtain a reasonable agreement with
experiments for the relaxation rate, the optical conductivity,
and the resistivity in broad regions of temperature and doping,
in particular, in the underdoped region with a strong AF
short-range order (SRO).
 \par
In Sec.~II we formulate a general theory of the optical
conductivity within the memory-function formalism. The
application of this  theory to the $t$--$J$ model is given in
Sec.~III. Numerical results and discussion are presented in
Sec.~IV. In Sec.~V we summarize our results.

\section{Memory-function theory}

In the linear response theory of Kubo \cite{Kubo57}, the
frequency dependent conductivity is defined by the
current--current relaxation function,
\begin{equation}
\sigma_{xx}(\omega) = {\frac{i}{V}}\, (\!( J_{x} | J_{x}
)\!)_{\omega} = {\frac{1}{V}} \int_{0}^{\infty}dt {\rm e}^{i
\omega t} (J_{x}(t), J_{x}),
 \label{g1}
\end{equation}
where $V$ is the volume of the system. Here, the Kubo--Mori
scalar product
\begin{equation}
   (A(t), B) =\int_{0}^{\beta}d \lambda
   \langle A(t-\imath \lambda) B \rangle
\label{g2}
\end{equation}
for the operators in the Heisenberg representation, $ A(t)=$  $
\exp (i Ht) A\exp (-i Ht)$, is introduced . $\langle AB \rangle$
denotes the equilibrium statistical average for a system with the
Hamiltonian $H$ (here $\beta = 1/T, $ $\hbar = k_{B} = 1$). The
real part of the conductivity (\ref{g1}) obeys the sum rule
\begin{equation}
  \int_{0}^{\infty}d\omega {\rm Re} \sigma_{xx}(\omega)
   = \frac{\pi}{2V}\chi_{0} =
   \frac{ i \pi}{2V} \langle [J_{x}, P_{x}]\rangle .
\label{g3}
\end{equation}
Here  $\, { P_{x}} = e\sum_{i} { R}^{x}_{i} N_{i} \,$ is the
polarization operator.  $ R_{i}^{x}$ is the $x$-component of the
lattice vector pointing to site $i$, $e$ is the electron charge,
and $N_i$ is the number operator. The current operator is defined
by the time derivative of the polarization operator:
 $\, J_{x}(t) = d P_{x}(t)/dt  \equiv  \dot{P}_{x}(t) = -
i[ P_{x}, H ] $. The static current-current susceptibility
$\chi_{0} = ( J_{x}, J_{x} ) $ is related to the effective number
of charge carriers participating in the absorption,
\begin{equation}
N_{\rm eff} = \frac{2 m v_{0}}{\pi {\rm e}^{2}} \,
\int\limits_{0}^{\infty} d\omega {\rm Re} \sigma_{xx}(\omega)
 = \frac{ m }{ {\rm e}^{2} N}\chi_{0} \,,
 \label{g3a}
\end{equation}
where $v_{0} = V/N$ is the unit cell volume and $m$ is the free
electron mass. The sum rule  (\ref{g3}) is frequently written in
terms of the plasma frequency $\omega_{\rm pl}$ defined by:
$\omega_{\rm pl}^2 = 4\pi \chi_{0}/ V = \omega_{0,\rm pl}^2 \,
N_{\rm eff}$, where $\omega_{0,\rm pl}^2 = 4\pi N  e^2/m V $.
\par
To calculate the conductivity (\ref{g1}), it is convenient to
employ the memory-function approach of Mori~\cite{Mori65} by
introducing the memory function $ M(\omega) $ for the relaxation
function~\cite{Goetze72},
\begin{equation}
\Phi(\omega) \equiv  (\!( J_{x} | J_{x} )\!)_{\omega} =
               \frac{\chi_{0}}{\omega + M(\omega)}.
\label{g4}
\end{equation}
From the equations of motion for the  relaxation function
 $\Phi(t-t')= (\!( J_{x}(t) | J_{x}(t'))\!) $  the memory
function is determined by (see Appendix A)
\begin{equation}
M(\omega) =  (\!(F_{x}|F_{x})\!)_{\omega}^{(\rm proper)}
(1/\chi_{0}),
 \label{g5}
\end{equation}
where  $F_{x} = i \dot{J}_{x} = [ J_{x} , H ]$ is the force
operator. The definition of the memory function (\ref{g5}) as the
``proper'' part of the force-force relaxation function   is
equivalent to the introduction of the projected Liouvillian
superoperator for the memory function in the original Mori
technique~\cite{Mori65}.
\par
Using Eq.~(\ref{g4}), the frequency dependent conductivity
(\ref{g1}) can be written in the form of the generalized Drude
law:
\begin{equation}
\sigma_{xx}(\omega)  \equiv \sigma(\omega) = \frac{\omega_{\rm
pl}^2}{4 \pi} \quad
      {\frac{m}{\widetilde{m}(\omega)}} \quad
      {\frac{1}{\widetilde{\Gamma} (\omega) - i \omega}},
\label{g6}
\end{equation}
where the effective optical mass and the relaxation rate are
given by
\begin{eqnarray}
 {\frac{\widetilde{m}(\omega)} {m}}& = & 1 + \lambda(\omega),
  \quad
\widetilde{\Gamma} (\omega) =
\frac{\Gamma(\omega)}{1+\lambda(\omega)} ,
 \label{g7} \\
\lambda(\omega) &= & M{'}(\omega)/\omega, \quad
          \Gamma(\omega) = M{''}(\omega).
\label{g8}
\end{eqnarray}
Here the real   and imaginary  parts of the retarded memory
function $\, M(\omega + i 0^+) = M{'}(\omega) + i M{''}(\omega)$
are introduced . They are coupled by the dispersion relation:
\begin{equation}
M'(\omega) = \frac{1}{\pi}\int_{-\infty}^{\infty} d\omega' \;
\frac {M''(\omega') }{\omega'-\omega}.
 \label{g8a}
 \end{equation}
Both the real and imaginary parts of the memory function can be
directly related to experimental data for the  inverse
conductivity (\ref{g6})~\cite{Basov05}:
\begin{equation}
 \Gamma(\omega)= \frac{\omega_{\rm pl}^{2}}{ 4\pi}\, {\rm Re}
\frac{1}{ \sigma(\omega)}, \quad
 1+ \lambda(\omega) = - \frac{\omega_{\rm pl}^{2}}{4\pi \omega}\,
 {\rm Im} \frac{1}{ \sigma(\omega)}.
 \label{g9}
\end{equation}
Below we calculate these functions  for the $t$--$J$ model.

\section{Relaxation rate}

We consider the  $t$--$J$ model on the square lattice which  in
the conventional notation reads:~\cite{Anderson87,Zhang88}
\begin{eqnarray}
H &= & H_t + H_J =  - \sum_{i \neq j,\sigma}\,t_{ij} \tilde
a_{i\sigma}^{+} \tilde a_{j\sigma} -
    \mu \sum_{i }\, N_i
 \nonumber \\
& + & \frac{1}{2} \sum_{i \neq j,\sigma} J_{ij} ({\bf S}_{i}{\bf
S}_{j} - \frac {1}{4} N_{i} N_{j}), \; \label{r1a}
\end{eqnarray}
where  $t_{ij}$ is the hopping integral and $J_{ij}$ is the AF
exchange interaction. Here $\tilde
a_{i\sigma}^{+}=a_{i\sigma}^{+}(1-n_{i\bar{\sigma}})$ is the
projected electron operator with spin $\sigma/2 = \pm 1/2, \;(
\bar{\sigma} = - \sigma) $ on the lattice site $i$, $N_i =
\sum_{\sigma}\tilde a_{i\sigma}^{+}\tilde a_{i\sigma} $ is the
number operator, and $\, S_{i}^{\alpha} = (1/2)\sum_{s,s'}\tilde
a_{is}^{+}\sigma^{\alpha}_{s,s'}\tilde a_{is'}\,$ is  the
$\alpha$-component of the spin operator ($\sigma^{\alpha}_{s,s'}$
are Pauli matrices). The chemical potential $\mu$ is determined
from the equation for the average electron occupation number
$\langle N_{i} \rangle = 1- \delta $, where $\delta$ is the hole
concentration.
\par
To take into account  the projected character of electron
operators  we employ the Hubbard operator (HO)
technique~\cite{Hubbard65,Izyumov89}. The HO
$\,X_{i}^{\alpha\beta}=|i,\alpha\rangle\langle i,\beta| \,$
describes the transition from the state $|i,\beta\rangle$ to the
state $|i,\alpha\rangle$ at the site $i$, where  $ \alpha$ and
$\beta$ denote three possible states: an empty state $(\alpha,
\beta =0) $ and a singly occupied state $(\alpha, \beta =
\sigma)$. The completeness relation $ X_{i}^{00} + \sum_{\sigma}
X_{i}^{\sigma\sigma} = 1$  rigorously preserves  the constraint
of no double-occupancy of any lattice site. From the
multiplication rule $\, X_{i}^{\alpha\beta} X_{i}^{\gamma\delta}
= \delta_{\beta\gamma} X_{i}^{\alpha\delta} \,$ follow the
commutation relations for the HOs:
\begin{equation}
 [X_{i}^{\alpha\beta}, X_{j}^{\gamma\delta}]_{\pm}=
\delta_{ij}\left(\delta_{\beta\gamma}X_{i}^{\alpha\delta}\pm
\delta_{\delta\alpha}X_{i}^{\gamma\beta}\right) .
 \label{r2a}
\end{equation}
The upper sign refers to the   Fermi-like operators creating
($X_{i}^{\sigma 0}$) or annihilating ($X_{j}^{0\sigma}$)
electrons, while the lower sign refers to the Bose-like
operators, such as the number or spin operators:
\begin{equation}
N_{i}= \sum_{\sigma} X_{i}^{\sigma \sigma}, \;  S_{i}^{z} =
\frac{1}{2}\sum_{\sigma} \sigma\, X_{i}^{\sigma\sigma},\;
S_{i}^{\sigma} = X_{i}^{\sigma \bar{\sigma}}.
 \label{r2}
\end{equation}
The commutation relations result in  {\it the kinematic
interaction} for  HOs (see Eq.~(\ref{B2})). Note that the term
``kinematic interaction'' was introduced by Dyson~\cite{Dyson56}
for  spin operators.
\par
Using the HO representation for $\tilde a_{i\sigma}^{+} =
X_{i}^{\sigma 0} \; $,  $\tilde a_{j\sigma}= X_{j}^{0\sigma}$,
and Eq.~(\ref{r2})  we  write the Hamiltonian (\ref{r1a}) in the
form:
\begin{eqnarray}
H &=&  - \sum_{i \neq j,\sigma}t_{ij}\, X_{i}^{\sigma
0}X_{j}^{0\sigma}
 - \mu \sum_{i \sigma} X_{i}^{\sigma \sigma}
\nonumber \\
 &  + &\frac{1}{4} \sum_{i \neq j,\sigma} J_{ij}
\left(X_i^{\sigma\bar{\sigma}}X_j^{\bar{\sigma}\sigma}  -
   X_i^{\sigma\sigma}X_j^{\bar{\sigma}\bar{\sigma}}\right) \, .
\label{r1}
\end{eqnarray}
\par
The relaxation rate $\Gamma(\omega) = M{''}(\omega)$ is
calculated by Eq.~(\ref{g5})  in the mode-coupling approximation
(MCA) as described  in Appendix~B. In this approximation we
obtain:
\begin{eqnarray}
&&\Gamma(\omega) ={\frac{(e^{\beta\omega} -1)}{\chi_{0}\,
\omega}} \, {\frac {2 \pi\,{\rm e}^2}{N}} \sum_{\bf k, q} \int \!
\!\int \! \! \int_{-\infty}^{\infty} d\omega_{1}
d\omega_{2}d\omega_{3}
\nonumber \\
&& \times
 n(\omega_{1})[1-n(\omega_{2})]N(\omega_{3})\,
 \delta(\omega_{2}-\omega_{1}- \omega_{3} +\omega)
\nonumber  \\
&& \times   g^{2}_{x} ({\bf k, k-q}) \chi_{cs}''({\bf
q},\omega_{3}) A({\bf k},\omega_{1})A({\bf k-q},\omega_{2})
 \label{r3} \; ,
\end{eqnarray}
where  $n(\omega)= (\exp \beta\omega  + 1)^{-1}$  and $N(\omega)
= (\exp\beta\omega - 1)^{-1}$. The momentum dependent (transport)
vertex is given by
\begin{eqnarray}
g_{x}({\bf k,k-q}) &= &v_{x}({\bf k})\, t({\bf k-q})- v_{x}({\bf
k-q})\,t({\bf k})
  \nonumber \\
&- &  J({\bf q})/2\, [v_{x}({\bf k}) -  v_{x}({\bf k-q})],
 \label{r4}
\end{eqnarray}
where $ t({\bf k})$ and $J({\bf q})$ are the Fourier transforms
of the  hopping integral and  the exchange interaction, and $\,
v_{x}({\bf k}) = - \partial {t({\bf k})}/ \partial {k_{x}} $ is
the electron velocity (see Appendix B).
 The spectral function for the charge-spin excitations
 $\chi_{cs}'' ({\bf q},\omega) = (1/\pi) {\rm Im}\,
 \chi_{cs}({\bf q},\omega) $ is defined by  the corresponding
commutator Green functions (GFs),
\begin{equation}
\chi_{cs}({\bf q},\omega) =
 - (1/4)\langle \!\langle N_{\bf q}| N_{\bf -q}
 \rangle \! \rangle_{\omega}
   - \langle \!\langle {\bf S}_{\bf q}|
{\bf S}_{\bf -q}\rangle \! \rangle_{\omega},
 \label{r5}
\end{equation}
where we used Zubarev's notation  \cite{Zubarev60} for the
retarded two-time GFs. The spectral function of electronic
excitations is defined by the imaginary part of the
anticommutator electronic GF,
\begin{equation}
  A({\bf k},\omega) =  - (1/\pi)\;
   {\rm Im} \langle\!\langle X^{0\sigma}_{\bf k}\mid
X^{\sigma 0}_{\bf k}\rangle\!\rangle_{\omega} \; .
 \label{r6}
 \end{equation}
The static current-current  susceptibility $\chi_{0}$ is connected
with the effective number of charge carriers (\ref{g3a}) which for
the  $t$--$J$ model reads
\begin{eqnarray}
N_{\rm eff} & = &
 \frac{ m }{ N} \sum_{i, j, \sigma} (R_{i}^{x} - R_{j}^{x})^{2}
  t_{ij}  \langle X^{\sigma 0}_{i}  X^{0\sigma }_{j} \rangle
\nonumber\\
 & = &
 -\frac{ m }{ N}  \sum_{{\bf k},\sigma}
  {\frac {\partial^2 t({\bf k})} {\partial k_{x}^2} }
  \langle X^{\sigma 0}_{{\bf k}}
X^{0\sigma }_{{\bf k}} \rangle.
 \label{r7}
\end{eqnarray}
For the  $t$--$J$ model with the nearest-neighbor hopping only, $
(R_{i}^{x} - R_{j}^{x})^2 = a^2 $, where $a$ is the lattice
parameter, the effective number of carriers is related to the
averaged kinetic energy, $ N_{\rm eff} = (a^2 m/ N) \langle -
H_t\rangle  $ . This relation is often used in the calculation of
the charge stiffness (Drude weight) in finite-cluster studies
(see, e.g., Refs.~\cite{Stephan92,Dagotto94}).

\section{Results and discussion}

\subsection{Spectral functions}

In numerical calculations  we  have to use models for  the
charge-spin susceptibility (\ref{r5}) and the one-electron
spectral function (\ref{r6}). The spin-excitation contribution in
Eq.~(\ref{r5}) is described by the spectral function
$\chi_s''({\bf q}, \omega)= (3/2)\chi_{\pm}''({\bf q}, \omega)$,
where $\chi_{\pm}''({\bf q}, \omega) =- (1/\pi){\rm
Im}\,\langle\!\langle S^+_{\bf q}|S^-_{-\bf
q}\rangle\!\rangle_{\omega}$. For the latter we use the function
calculated in Ref.~\cite{{Vladimirov09}} for the $t$-$J$ model,
\begin{eqnarray}
\chi_{\pm}''({\bf q}, \omega)  =  \frac{- \omega \,
\Sigma_s{''}({\bf q})\;( m({\bf q})/ \pi)} {[\omega^2 -
\omega_{\bf q}^2 - \omega \, \Sigma_s{'}({\bf q},\omega)]^2  +
[\omega \,  \Sigma_s{''}({\bf q})]^2} .
 \label{n3}
\end{eqnarray}
Here, the spectrum of spin excitations in the generalized
mean-field approximation  $ \omega_{\bf q}$ determines the static
spin susceptibility $\chi_{\bf q} =  m({\bf q})/\omega_{\bf q}^2
$ with $m({\bf q})=\langle [i\dot{S}^{+}_{\bf q}, S_{-\bf
q}^{-}]\rangle$,  where $i\dot{S}^{+}_{\bf q} = [{S}^{+}_{\bf q}
, H]$ . The self-energy $\, \Sigma_s({\bf q},\omega)=
\Sigma_s{'}({\bf q},\omega) + i \Sigma_s{''}({\bf q},\omega)$,
where $\,\Sigma_s{'}({\bf q},\omega)$ and $ \Sigma_s{''}({\bf
q},\omega)$ are the real and the imaginary parts, respectively,
is determined by the many-particle relaxation function $\,
\Sigma_s({\bf q},\omega)=(1/m({\bf q}))(\!( - \ddot{S}_{\bf
q}^{+}| -\ddot{S}_{-\bf q}^{-})\!)_{\omega}^{\rm proper} $
calculated in MCA (see Refs.~\cite{Vladimirov09,Vladimirov11}).
Taking into account that the main contribution to the relaxation
rate (\ref{r3})   from the spectral function (\ref{n3}) is given
by frequencies close to the renormalized spin-excitation frequency
$\,\widetilde{\omega}_{\bf q} = [\omega_{\bf q}^2 +
\widetilde{\omega}_{\bf q} \, \Sigma'_s({\bf
q},\widetilde{\omega}_{\bf q})]^{1/2} \,$, we approximate the
damping of spin excitations by the function $\Sigma_s{''}({\bf
q})= \Sigma_s{''}({\bf q},\omega = \widetilde{\omega}_{\bf q})$.
\par
To calculate the contribution to the relaxation rate (\ref{r3})
from charge (density) excitations in Eq.~(\ref{r5}), we use the
spectral function calculated in Ref.~\cite{Jackeli99} for the
$t$--$J$ model. Our results show that charge excitations give the
main contribution in the region of high energies, $ \omega \sim
3t$, which, however, is several times weaker than the
spin-excitation contribution and, therefore, can be safely
ignored. The different energy scales for  spin excitations,
$\omega \sim J$, and density excitations, $\omega \sim t$, were
also found in an exact diagonalization study of the $t$--$J$
model~\cite{Eder95}.
\par
The self-consistent solution of the system of equations for the
spectral function (\ref{r6}) and the single-electron  self-energy
in Ref.~\cite{Plakida99} has shown that close to the Fermi energy
there  appear well-defined quasiparticle excitations. This result
permits to approximate the spectral function by the expression
\begin{equation}
 A({\bf k},\omega) = Q\, \delta (\omega -
 \tilde{\varepsilon}_{{\bf k}}),
 \label{n2}
\end{equation}
where $Q = 1-n/2$ is the  spectral weight for electronic
excitations in the  $t$--$J$ model. To model a realistic
electronic spectrum  which crosses the AF  Brillouin zone (BZ),
as observed in ARPES experiments  (see, e.g.,
Ref.~\cite{Kordyuk05}), we consider the model dispersion
\begin{equation}
\tilde{\varepsilon}_{{\bf k}} = - 4 Q[ t\,\alpha_1 \gamma({\bf
k}) + t' \alpha_2 \,\gamma'({\bf k}) +t'' \alpha_2
\,\gamma''({\bf k})] - \mu ,
 \label{n4}
\end{equation}
where  $t $ and $ t'= 0.1 t, \, t''=  0.2 t$ are the hopping
parameters for the nearest and further-distant neighbors,
respectively, and $\gamma({\bf k})= (1/2)(\cos a k_x +\cos a
k_y), \,  \gamma'({\bf k}) = \,\cos a k_x \cos a k_y\, $ and
$\gamma '' ({\bf k})= (1/2)(\cos 2a k_x +\cos 2a k_y) $. The
kinematic interaction for the HOs results in a renormalization
of  the spectrum  (\ref{n4}) determined by the parameters $\,
\alpha_1 = [ 1 + {C_{1}}/{Q^2}]\,$ and $\, \alpha_2 = [ 1 +
{C_{2}}/{Q^2}]\,$, where $\, C_{1} = \langle {\bf S}_{\bf i}{\bf
S}_{{\bf i}\pm {\bf a}_{x}/{\bf a}_{y}} \rangle$ and $ C_{2} =
\langle {\bf S}_{\bf i}{\bf S}_{{\bf i}\pm {\bf a}_{x}\pm {\bf
a}_{y}} \rangle \approx \langle {\bf S}_{\bf i}{\bf S}_{{\bf
i}\pm 2{\bf a}_{x}/2{\bf a}_{y}} \rangle\,$  are the spin
correlation functions for the nearest and the second neighbors,
respectively (see Ref.~\cite{Plakida99}). With increasing doping,
the effective bandwidth $\widetilde{W}$ of the dispersion
(\ref{n4}) increases due to the decrease of AF SRO described by
the doping dependence of the spin correlation functions in the
renormalization parameters $\alpha_1, \alpha_2$. In particular,
for $\delta = 0.09\, (0.2)$ at $T = 0$  we have $\widetilde{W} =
1.14\,t \, (2.78 \,t)\,$ in comparison with the unrenormalized
bandwidth $W = 8\, t\, Q = 4.36 \,t\, (4.8 \,t)$.
\par
\begin{figure}[ht!]
\includegraphics[width=0.35\textwidth]{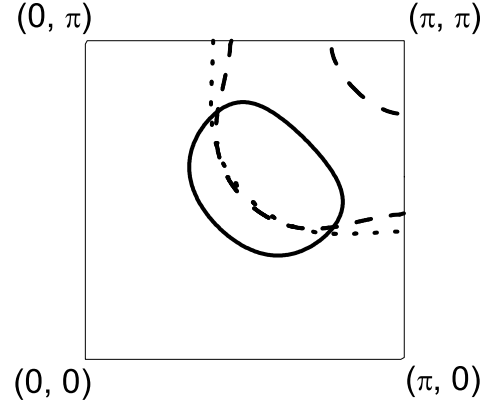}
\caption[]{Fermi surface at $\delta = 0.09$ (solid line), $0.16$
(dashed line), and $0.2$ (dots).}
 \label{FS}
\end{figure}

The Fermi surface (FS) determined by the equation
$\tilde{\varepsilon}_{\bf k_{\rm F}} =0 $ is shown in
Fig.~\ref{FS} for various doping. The renormalization of the
spectrum induced by the AF SRO provides a FS with hole pockets at
low doping  which is equivalent to  a pseudogap in the $(\pm
\pi,0)$ and $ (0, \pm \pi)$ regions of the BZ. In the study of the
electronic spectrum and the FS of cuprates, a more accurate
calculation of the spectral function (\ref{n2}) including the
self-energy contribution was performed  as reported in
Refs.~\cite{Plakida99} and~\cite{Prelovsek02} for the $t$--$J$
model and in Ref.~\cite{Plakida07} for the two-subband Hubbard
model. In particular, in Ref.~\cite{Plakida07} the spectral
function on the FS close to the $(\pi,\pi)$-point of the BZ has a
weak intensity resulting in an arc-type FS as observed in ARPES
experiments. By taking into account that the spectral function of
spin excitations (\ref{n3}) is peaked at the AF wave vector ${\bf
Q} = (\pi,\pi)$  and is very broad in other parts of the BZ (see
Ref.~\cite{Vladimirov11}), in the calculation of the relaxation
rate (\ref{r3}), only those parts of the FS are important which
are coupled by the AF wave vector  ${\bf Q}$. Therefore, the
parts of the FS in Fig.~\ref{FS} far away from the AF BZ, in
particular the back side of the hole-pocket near $(\pi/2,\pi/2)$
of the BZ, give small contributions at the integration over the
BZ in Eq.~(\ref{r3}).  This reasoning justifies the quasiparticle
approximation (\ref{n2}) used in the calculation of the
relaxation rate.
\par
The  effective number of charge carriers  (\ref{r7})  is
convenient to write in the form
\begin{eqnarray}
&&  N_{\rm eff}(\delta) = \eta K(\delta) = \eta \frac{Q}{N}
\sum_{{\bf k},\sigma}\, n(\tilde{\varepsilon}_{\bf k})\,[\,\cos
(a k_x)
\label{n6}\\
 & + &  2(t'/ t)\cos (a k_x) \cos (a k_y)
 + 4(t''/t)\cos (2 a k_x)].
\nonumber
\end{eqnarray}
For the spectral function (\ref{n2}) the average electron
occupation number is given by $\, \langle X^{\sigma 0}_{{\bf k}}
X^{0\sigma }_{{\bf k}} \rangle  = Q\, n(\tilde{\varepsilon}_{\bf
k})\,$. The prefactor  $\eta = 2 m a^2 t = t \, (p^2/2m)^{-1}$ is
a dimensionless ratio of the hopping parameter $t$ to the kinetic
energy of  an electron with the momentum $ \, p $. In particular,
for $a = 3.8$~\AA $\;$ and $t = 0.4 $~eV we have $\,\eta = 3.79
\; t\,[{\rm eV}] =1.52$.
\par
In these  approximations,  the relaxation rate (\ref{r3}) is
determined by the expression
\begin{eqnarray}
 \Gamma(\omega)& = & t \,\frac{\pi(e^{\beta\omega} -1)}
 {  \omega \,K(\delta)}
 \, \frac {3 t^2 \,Q^2}{2 N^2}\, \sum_{\bf k, q}
 \int_{-\infty}^{\infty} d\omega'\, \chi_{\pm}''({\bf q}, \omega')
 \nonumber \\
&& \times   \widetilde{g}^{2}_{x} ({\bf k,k-q}) \,
\delta(\tilde{\varepsilon}_{{\bf k}}- \tilde{\varepsilon}_{{\bf
k-q}} + \omega' - \omega)
 \nonumber\\
&& \times  N(\omega')\, n(\tilde{\varepsilon}_{{\bf k}})
 \, [1-n(\tilde{\varepsilon}_{\bf k - q})]  ,
 \label{n5}
\end{eqnarray}
where, using  Eq.~(\ref{r4}), the dimensionless transport vertex
$\widetilde{g}_{x} ({\bf k, k-q}) = (1/ a t^2 )\,g_{x} ({\bf k,
k-q})\,$ is introduced.
\par
The real part of the conductivity (\ref{g6}) may be written as
\begin{eqnarray}
{\rm Re}\,\sigma(\omega) =
      A\, \widetilde{\sigma}(\omega)\equiv A
      \frac{N_{\rm eff}\,t\,\Gamma(\omega)}{[\omega + M'(\omega)]^2
      + [\Gamma(\omega)]^2 } ,
      \label{n7}
\end{eqnarray}
where $ A =  \omega_{0 {\rm pl}}^2/ (4\pi t)= e^2/ (m v_0 t) $.
\par
The real part of the memory function $ M'(\omega)= \omega \,
\lambda(\omega)$ is calculated by the dispersion relation
(\ref{g8a}) using the relaxation rate (\ref{n5}). This enables us
to calculate  the effective optical mass (\ref{g7}),
$\widetilde{m}/m = 1 + \lambda(0) $.

In numerical calculations we take $J = 0.3\,t$ and  $\, t =
0.4$~eV  as an energy unit (0.4~eV  = 3226~cm$^{-1}$ = 4640~K).
The results for the relaxation rate and the optical conductivity
as functions of frequency, temperature, and hole doping are in a
good overall agreement with experiments. This will be detailed in
the following.

\subsection{Relaxation rate}

\begin{figure}[ht!]
\includegraphics[width=0.35\textwidth]{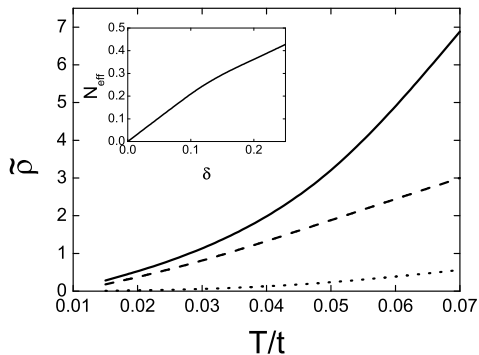}
\caption[]{Resistivity  $\widetilde{\rho}(T) =
1/\widetilde{\sigma}(0, T )$ for doping $\delta = 0.09$ (solid) ,
$0.16$ (dashed), and $0.2$ (dotted). In the inset the effective
number of charge carriers  $N_{\rm eff}(\delta)$ at  $T = 0$ is
shown.}
 \label{figRN}
\end{figure}

At zero frequency, the relaxation rate is related to the
dimensionless electrical resistivity $\widetilde{\rho} =
1/\widetilde{\sigma}(0)$ by $\Gamma(0)/t = N_{\rm
eff}\widetilde{\rho}$. The temperature dependence of
$\widetilde{\rho}$ for $\delta = 0.09$, $0.16$, and $0.20$ is
shown in Fig.~\ref{figRN}. For a doping near and larger than the
optimal doping ($\delta = 0.16$), we obtain a nearly linear
temperature dependence, as is also observed in experiments (see,
e.g., Refs.~~\cite{Boebinger96,Ando04}).
\par
The effective number of charge carriers $N_{\rm eff} $ given by
Eq.~(\ref{n6}) is shown in the inset of Fig.~\ref{figRN}. It does
not reveal a notable temperature dependence and can be
approximated by the function $N_{\rm eff} \simeq 2 \,\delta$. It
is remarkable that $N_{\rm eff}$ increases more rapidly than the
hole doping. This result is in agreement with the in-plane
optical conductivity data on LSCO compounds
(Refs.~\cite{Uchida91,Uchida96}) which yields the effective
number of charge carriers $N_{\rm eff}(\omega)$ involved in
optical excitations up to the cut-off frequency $\omega$ (upper
limit of the integral in Eq.~(\ref{g3a})). $N_{\rm eff}$ was
found to be nearly proportional to $2 \delta$ for doping $\delta
< 0.15$, e.g., at $\delta = 0.1$, $N_{\rm eff}(\omega = 1.5\,
{\rm eV})\, = 0.19$ (Fig.~11 in Ref.~\cite{Uchida91}) and $N_{\rm
eff}(\omega = 2\, {\rm eV})\, = 0.26$  (Fig.~10 in
Ref.~\cite{Uchida96}).
\par
\begin{figure}[ht!]
\includegraphics[width=0.35\textwidth]{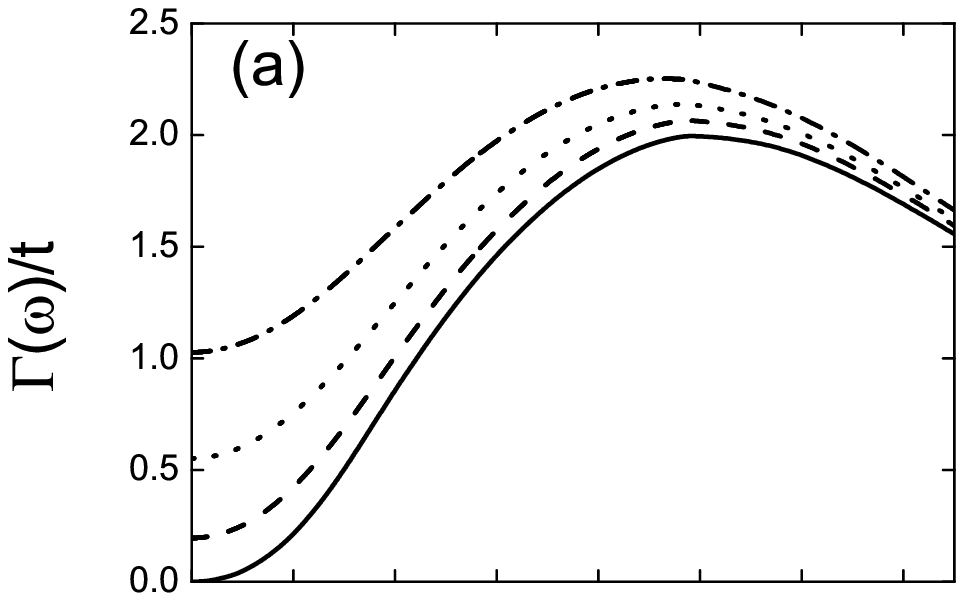}
\includegraphics[width=0.35\textwidth]{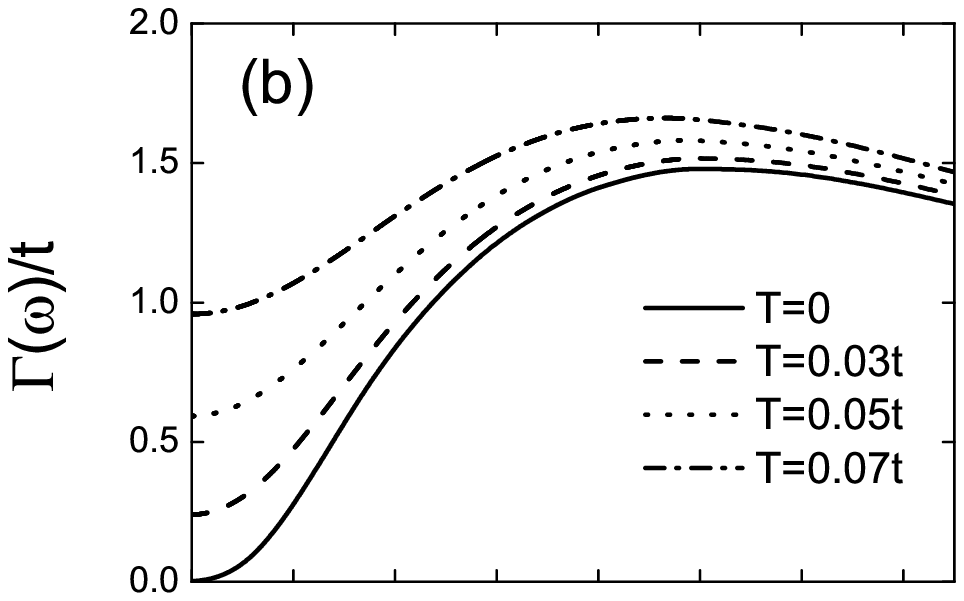}
\includegraphics[width=0.35\textwidth]{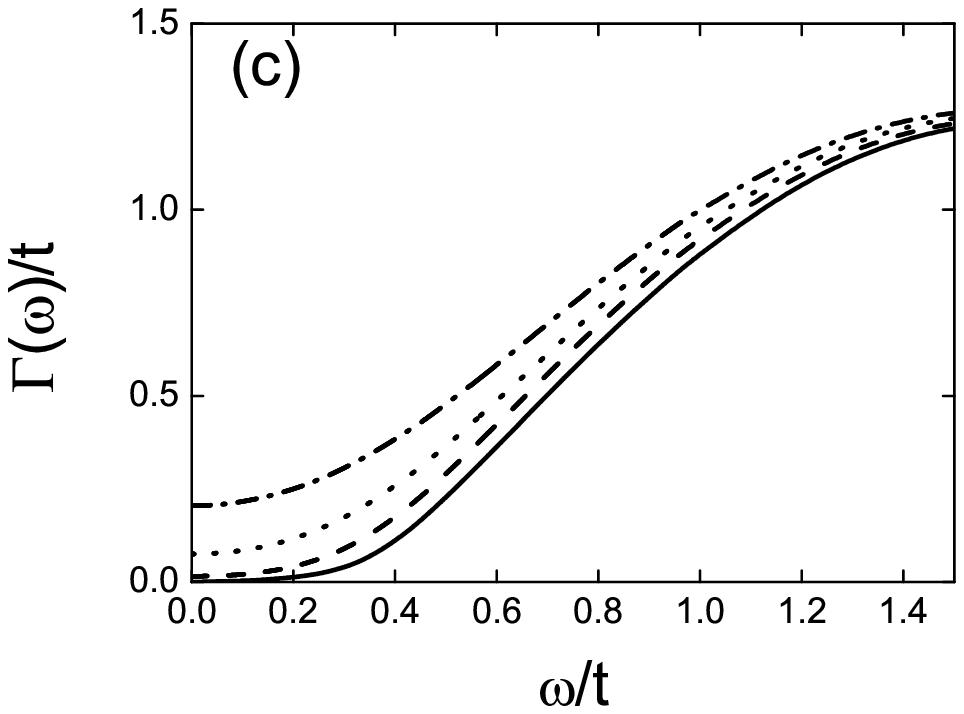}
\caption[]{ Temperature dependence of the relaxation rate $
\Gamma(\omega)$ for  (a) $\delta = 0.09$, (b) $0.16$, and (c)
$0.2$. Note the different scales. }
 \label{figG}
\end{figure}

The frequency dependence of the relaxation rate $\Gamma(\omega)$
(\ref{n5}) at different temperatures and doping is plotted in
Fig.~\ref{figG}. We obtain an increase  of $\Gamma(\omega)$ with
increasing temperature, which qualitatively agrees with
experiments. In the overdoped case, the relaxation rate decreases
resulting from the suppression of  spin fluctuations. The broad
maximum in the frequency dependence of $\Gamma(\omega)$, clearly
revealed at low doping in Fig.~\ref{figG}~(a), shifts to higher
frequencies with increasing doping.  The doping-dependent finite
effective bandwidth  $\widetilde{W}$  limits the highest
frequency for the relaxation, $\omega \leq 2 \widetilde{W} $, so
that at very high frequencies, $\Gamma(\omega)$ vanishes
according to $ \Gamma(\omega) \propto 1/\omega \rightarrow 0$.
Let us point out that a maximum in the relaxation rate is also
observed in experiments for the underdoped samples as, e.g., in
YBa$_2$Cu$_3$O$_{y}$ (YBCO$_y$) at $\omega \sim 2000$~cm$^{-1}$
for $y \lesssim 6.5$~\cite{Lee05}.
\par
\begin{figure}[ht!]
\includegraphics[width=0.35\textwidth]{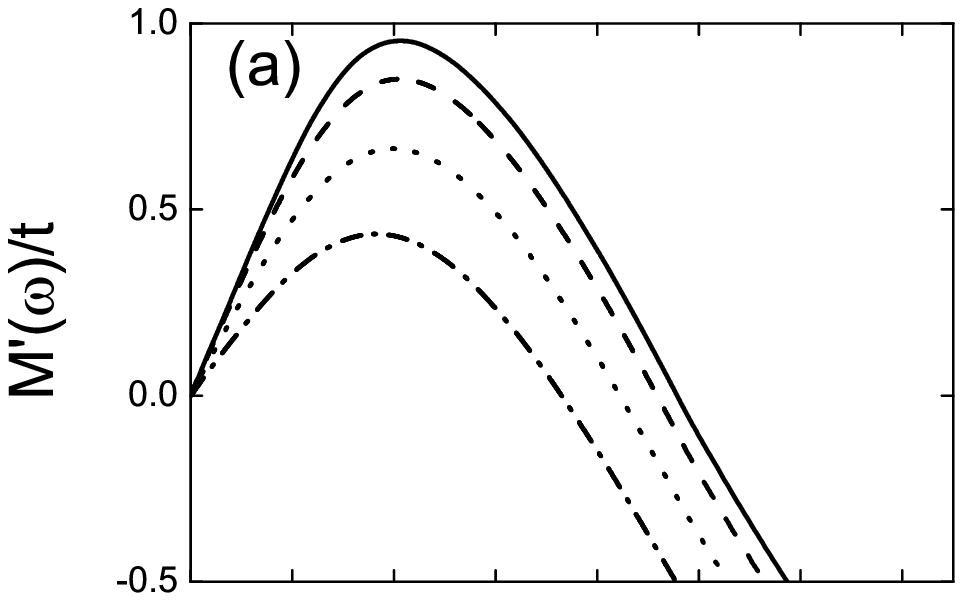}
\includegraphics[width=0.35\textwidth]{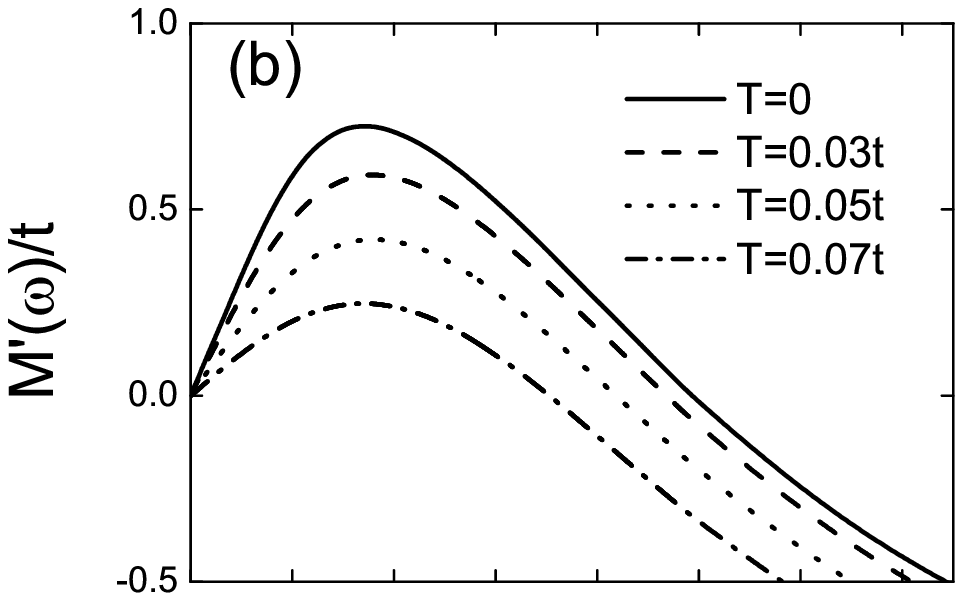}
\includegraphics[width=0.35\textwidth]{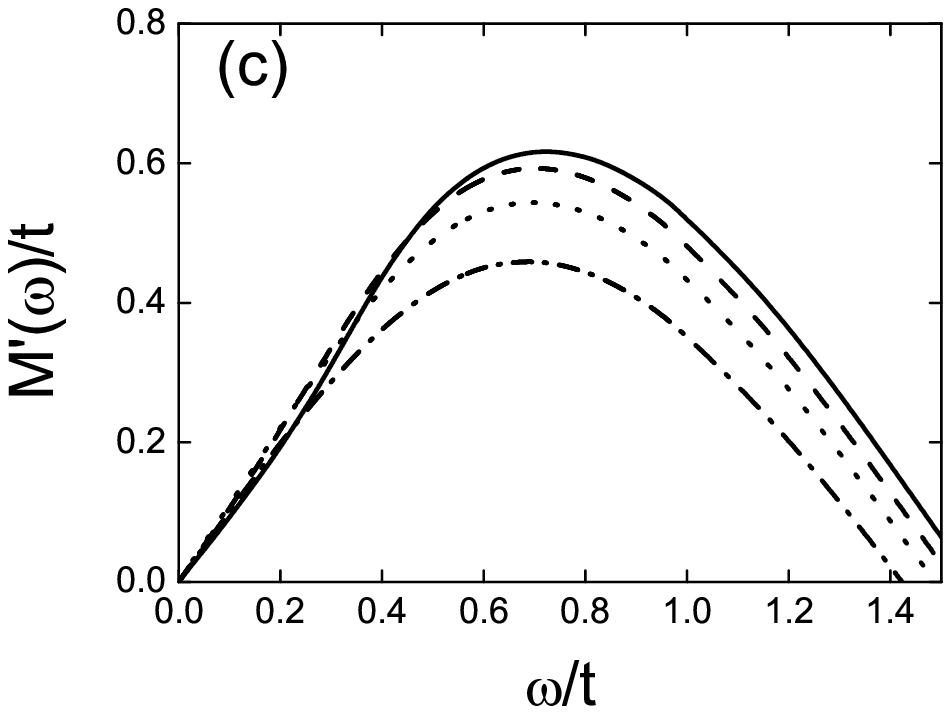}
\caption[]{Temperature dependence of the real part of the memory
function  $M'(\omega)$ for (a) $\delta = 0.09$, (b) $0.16$, and
(c) $0.2$. Note the different scales.}
 \label{figM}
\end{figure}

The real part of the memory function $ M'(\omega)$ shown in
Fig.~\ref{figM}  exhibits a maximum which height  decreases with
increasing temperature and doping.  But the energy of the peak
does not change with temperature as observed in experiments  (see,
e.g.,~\cite{Hwang06,Yang09,Heumen09}). In the underdoped case
($\delta = 0.09$), the temperature dependence of $ M'(\omega)$ is
very strong, as compared with the overdoped case ($\delta =
0.2$). This results from  the strong AF SRO at low doping that
strongly depends on temperature. With increasing doping, both the
SRO and the influence of temperature are weakened. Qualitatively,
the relaxation rate (Fig.~\ref{figG}) reveals the same trend.
\par
\begin{figure}[ht!]
\includegraphics[width=0.35\textwidth]{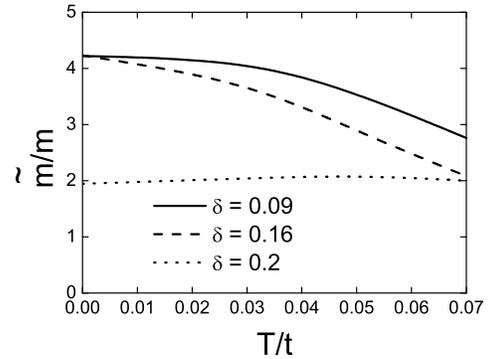}
\caption[]{Temperature dependence of the effective  optical mass
$\widetilde{m}/m  = 1 + \lambda(0)$ at various doping.}
 \label{figm}
\end{figure}

In Fig.~\ref{figm} we  show the temperature dependence of the
effective optical mass  at zero frequency, $\widetilde{m}/m  = 1
+ \lambda(0)$, at various  doping. At small doping, a strong
temperature dependence of $\widetilde{m}/m$  is observed, which
may be explained similarly as for  $M'(\omega)$.  For the
overdoped case,  the effective mass shows a weak renormalization,
$\widetilde{m}/m \sim 2$. In numerical studies of the one-hole
motion in small clusters, a weak renormalization of the optical
mass was deduced at high temperatures, $T \gtrsim 0.2 t$
(Ref.~\cite{Jaklic00}) which is in agreement with our results. In
early experiments, Ref.~\cite{Uchida91}, a large effective mass
renormalization was obtained in LSCO ranging from
$\widetilde{m}/m \simeq  25$ for $\delta = 0.1$ to
$\widetilde{m}/m =  16$ for $\delta = 0.2$. However, later on,
e.g., in Ref.~\cite{Padilla05}, a nearly doping independent modest
renormalization of the effective mass, $\widetilde{m}/m = 3 - 4$,
was observed in both LSCO and YBCO which is close to our finding.

\subsection{Optical conductivity}

\begin{figure}[ht!]
\includegraphics[width=0.35\textwidth]{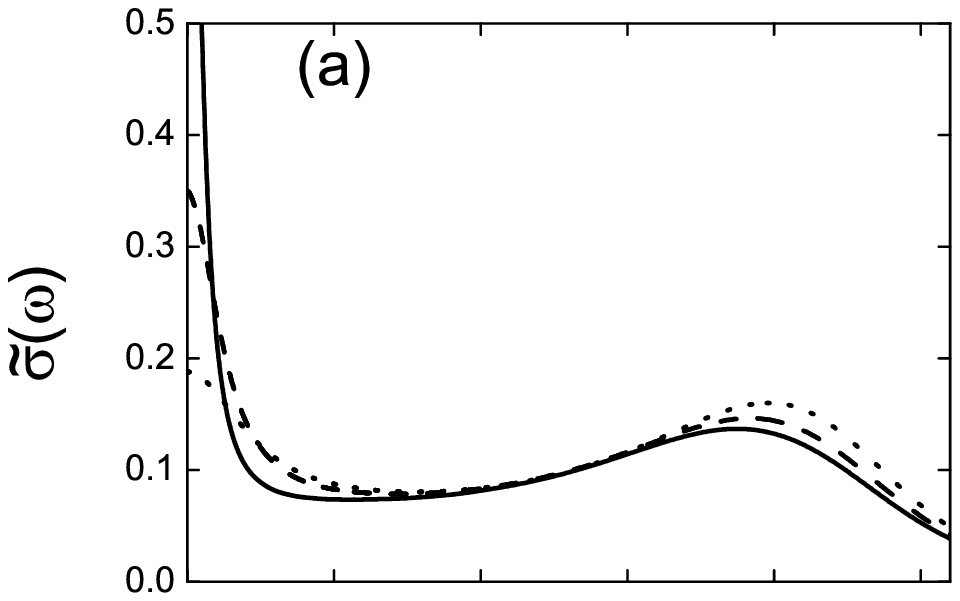}
\includegraphics[width=0.35\textwidth]{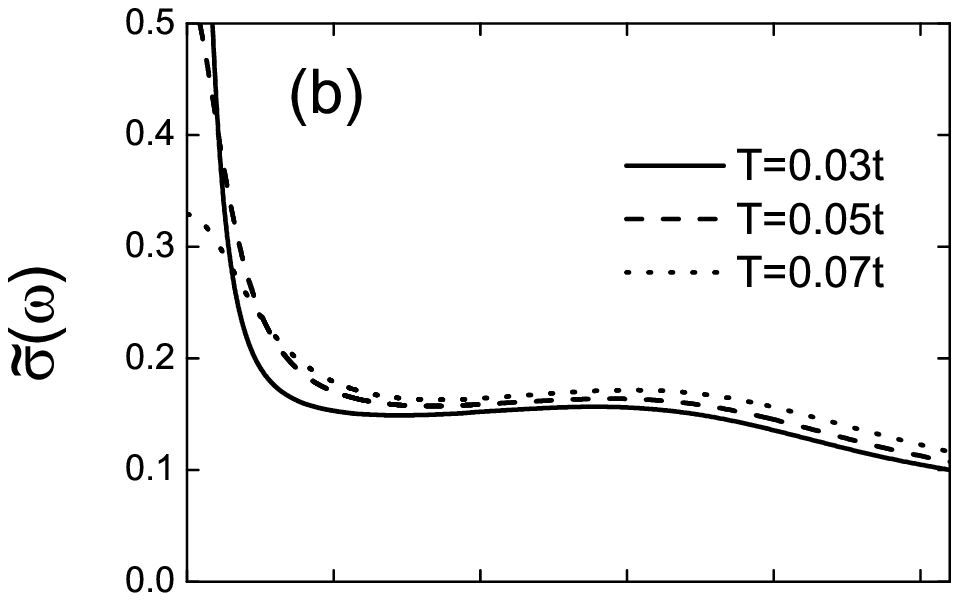}
\includegraphics[width=0.35\textwidth]{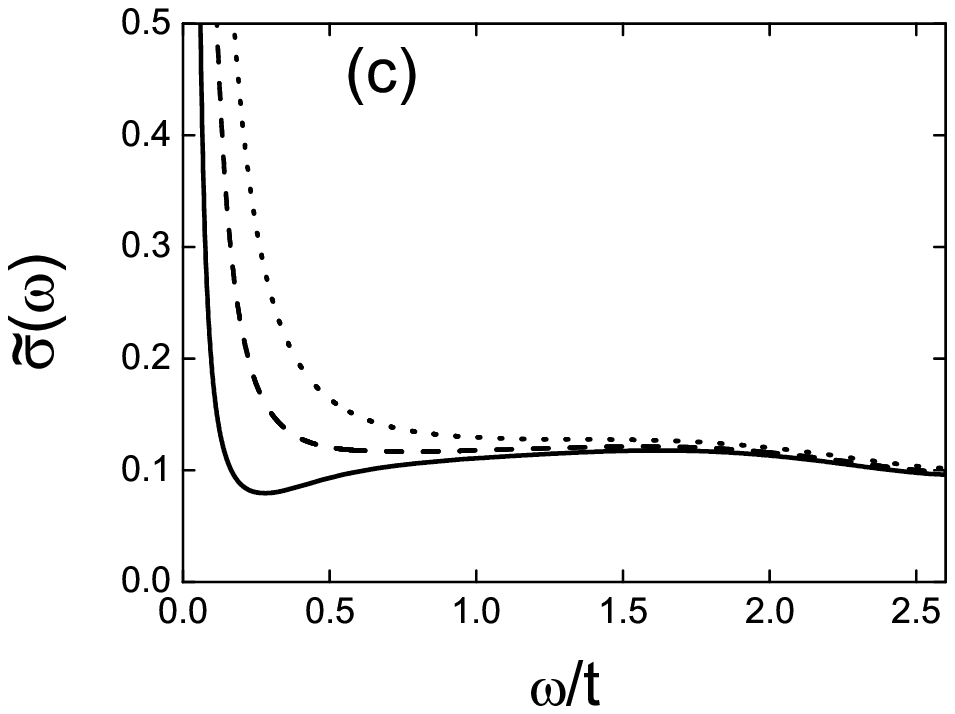}
\caption[]{Temperature dependence  of the optical conductivity
$\widetilde{\sigma}(\omega)$ for  (a) $\delta = 0.09$, (b)
$0.16$, and (c) $0.2$. }
 \label{figC}
\end{figure}

The frequency dependence of the conductivity (\ref{n7}) for
various temperatures and hole doping  is shown in Fig.~\ref{figC}.
The temperature and doping dependences of the conductivity show a
reasonable qualitative agreement with experiments (see, e.g.,
Ref.~\cite{Lee05}). At low frequencies, a large Drude peak is
found which significantly narrows at low temperatures. We also
obtain a broad and nearly temperature-independent MIR maximum at
$\omega \lesssim 2t \sim 6000$~cm$^{-1}$ which slightly shifts to
lower frequencies and becomes of lower intensity  with increasing
doping as observed in experiments (see, e.g., ~\cite{Padilla05}).
In our theory, the MIR absorption results from electron
interaction with  spin fluctuations which influence  the electron
scattering so that it decreases with increasing doping.

\subsection{Quantitative comparison to experiments}

\begin{figure}[ht!]
\includegraphics[width=0.34\textwidth]{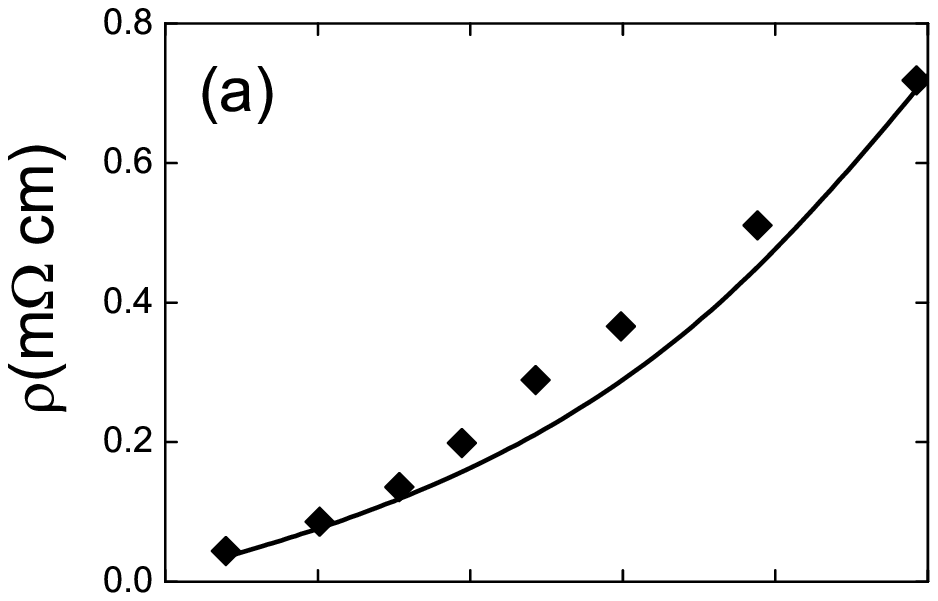}
\hspace*{1mm}
\includegraphics[width=0.35\textwidth]{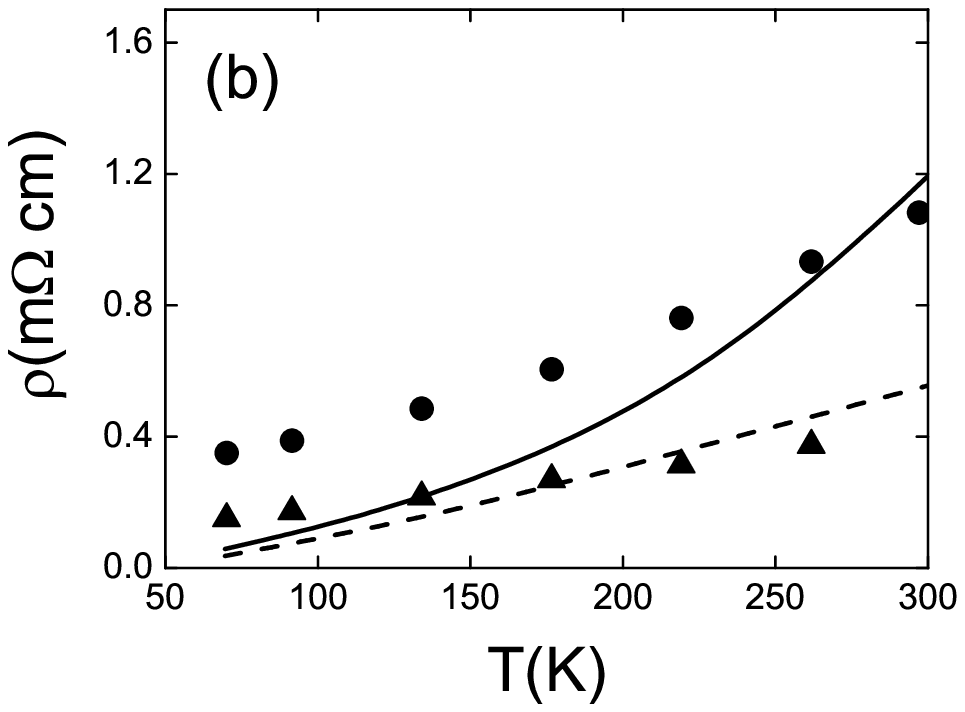}
\caption[]{Temperature  dependence of the resistivity $\rho(T)$
at (a) $\delta = 0.09$ in comparison with  the experimental data
for YBCO$_{6.5}$ from Ref.~\cite{Hwang06} shown by symbols, and
(b) at $\delta = 0.08$ (solid) and $\delta = 0.17$ (dashed) in
comparison with   the experimental data for  LSCO from
Ref.~\cite{Boebinger96} shown by symbols.}
 \label{figR_exp}
\end{figure}

Let us first compare the resistivity $ \rho(T) = (1/A)\,
\widetilde{\rho}$, where $A$ is given in Eq.~(\ref{n7}),  with
experimental data for the underdoped cuprate YBCO$_{6.5}$
($\delta = 0.09$)~\cite{Hwang06} shown  in
Fig.~\ref{figR_exp}~(a). Here  we use the value $ A = e^2/ (m v_0
t) = 8.00 \cdot 10^3[{\rm \Omega \cdot cm}]^{-1}$ taking
 $v_0  = 57.7$~(\AA)$^3$~\cite{Hwang06}. Without a fitting
procedure, we obtain a remarkably good agreement with experiment,
both in the absolute values of the resistivity and in its
temperature dependence. In Fig.~\ref{figR_exp}~(b) we compare the
resistivity  $ \rho(T)$ with experimental data on
LSCO~\cite{Boebinger96} for the underdoped $(\delta = 0.08)$ and
nearly optimally doped  $(\delta = 0.17)$ samples with our
results. The value of  $ A = 4.86 \cdot 10^3[{\rm \Omega \cdot
cm}]^{-1}$ is obtained  using $v_0  = a^2 d =95.3 $~(\AA)$^3$ ($a
=3.8$~\AA \, and $d =  6.6$~\AA \,). A reasonable agreement is
observed at high temperatures, while at low temperatures our
values are much smaller. An additional scattering mechanism,
e.g., impurity scattering,  should be invoked to explain the
experimental data. The comparison of our results with optimally
doped  and overdoped Tl-compounds~\cite{Ma06} shows the same
trend.
\par
Now we estimate the plasma frequency $\omega_{\rm pl} =
\left[N_{\rm eff}\right]^{1/2} \omega_{0, \rm pl}$. For the
optimally doped case, $\delta = 0.16$, we have $N_{\rm eff} =
0.3$ and $\omega_{\rm pl} = 0.55 \,\omega_{0, \rm pl}$ . For LSCO
 ($ \omega_{0, \rm pl} =  3.72$~eV) we get $\omega_{\rm pl}=
2.05$~eV  and for  YBCO ($\omega_{0, \rm pl}= 3.96$~eV),
$\omega_{\rm pl}= 2.18$~eV. These values are close to
experiments, while the LDA calculations in Ref.~\cite{Pickett89}
give  the somewhat larger value $\omega_{\rm pl} \approx 2.9$~eV.
For the underdoped YBCO$_{6.5}$ crystal, $\omega_{\rm pl} =
1.89$~eV~\cite{Hwang06}, while  for $\delta = 0.09$ we have
$N_{\rm eff} = 0.18$ and $\omega_{\rm pl} = 1.68$~eV which is
close to the experimental value.

\begin{figure}[ht!]
\includegraphics[width=0.35\textwidth]{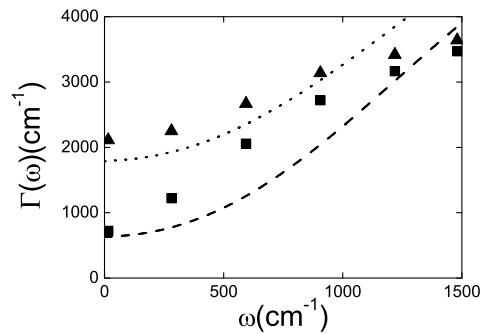}
\caption[]{Relaxation rate $ \Gamma(\omega)$ at  $\delta = 0.09$
for $T = 0.03t$ (dashed) and $T = 0.05t$ (dotted) in comparison
with experimental data for YBCO$_{6.5}$ (Ref.~\cite{Hwang06})
shown by symbols: squares for  $T = 147$~K and triangles  for $T =
244$~K.}
 \label{figG_YBCO}
\end{figure}
Finally, let us compare the relaxation rate $ \Gamma(\omega)$
with the optical data for YBCO$_{6.5}$ given in
Ref.~\cite{Hwang06}. In  the frequency region $\omega \leq
1500$~cm$^{-1}$ and for  the temperatures $T = 0.03t$  and $T =
0.05t$, which are close to the experimental values $T = 147$~K
and   $T = 244$~K, respectively, we get the results shown in
Fig.~\ref{figG_YBCO}. As we see, for the optical properties we
also obtain  a reasonable quantitative agreement of our theory
with experiments.

\subsection{Comparison with previous theoretical studies}

Various  methods have been used in theoretical studies of the
optical and dc conductivities in cuprates as discussed in Sec.~I.
Here we compare our results with previous studies to clarify what
kind of problems the latter has and how we have resolved some  of
them.
\par
One of the problems is how to explain a linear temperature
dependence of the resistivity in optimally doped cuprates in a
broad temperature range (see, e.g., Ref.~\cite{Friedman90} and a
discussion in Ref.~\cite{Ando04}). In early studies the
local-density-functional  theory was used in the calculation of
transport properties of cuprates  (for a review
see~\cite{Pickett89}). Calculations of the resistivity within the
relaxation rate approximation for electron scattering on phonons,
$\, (1/\tau_{\rm tr}) \propto \lambda_{\rm tr} \, T\,$, result in
a linear $T$-dependence over a broad temperature range, $\rho
\propto 1/(\tau_{\rm tr} \omega^{2}_{pl})$.  However, the absolute
values  of the resistivity prove to be several times smaller than
the experimental ones. This discrepancy could be removed by using
larger values of the transport EPI coupling constant
$\lambda_{\rm tr} = 1.5 -2 $ instead of the calculated value
$\lambda_{\rm tr} = 0.65 - 0.32\,$ for optimally doped LSCO and
YBCO, or by assuming the theoretical plasma frequencies
$\omega_{pl}$ to be smaller by a factor of two to three.
\par
To reconcile a weak transport EPI coupling $\lambda_{\rm tr} \sim
0.5$ and a strong EPI for quasiparticles,  $\lambda_{qp} \sim 2$,
needed to explain the high $T_c$ in cuprates, a model of a strong
forward scattering of electrons induced by electron correlations
was proposed~\cite{Zeyher96} which  enables to obtain
$\lambda_{\rm tr} \sim \lambda_{qp}/3$ (for a review
see~\cite{Maksimov10}, Sec.~III.B). This model was used in
Ref.~\cite{Varelogiannis98} to explain a linear $T$-dependence of
the resistivity in a broad region of $T$ where, however, the
extended van Hove singularities sufficiently close to the Fermi
energy  were assumed and fitting parameters  for EPI were
introduced.
\par
As shown in Fig.~\ref{figR_exp}~(b),  we obtain a linear
temperature dependence for the resistivity for a  nearly
optimally doped sample  in a reasonable agreement with experiment
on LSCO without using  fitting parameters. The values of the
plasma frequency  are also close to experiments. In our
memory-function theory  the transport relaxation rate
$\Gamma(\omega)$  and the quasiparticle self-energy are not
related in a simple way,  and therefore we can explain a
sufficiently weak scattering observed in the conductivity and a
strong superconducting pairing induced by the same electron
coupling to  spin fluctuations  (see, e.g.,~\cite{Plakida11}).
\par
To describe the normal state properties, such as  the resistivity
and OC, phenomenological spin-fermion models have been used. In
particular,  within a nearly antiferromagnetic Fermi-liquid
model~\cite{Millis90}  a reasonable agreement with experiments on
resistivity  for YBCO$_7$~\cite{Monthoux94} and on OC for
optimally doped  and overdoped cuprate
compounds~\cite{Stojkovic97} was obtained by a particular choice
of model parameters. In Ref.~\cite{Arfi92} the OC of YBCO$_{x},\;
(x = 6.3,\, 7)$ within the memory-function method was calculated.
To obtain an agreement with experiments, several models for the
spin susceptibility have to be considered. A detailed qualitative
discussion of the OC behavior  at various frequencies and
temperatures within the spin–-fermion model  was performed in
Ref.~\cite{Abanov03}. Contrary to these phenomenological
approaches,  we  obtain a fair agreement with experiments for
various doping and temperatures within the microscopic theory for
the spin-fluctuation susceptibility without fitting  coupling
parameters.
\par
There are several studies of the OC within the Hubbard model in
the limit of weak correlations, where the insulating AF state
emerges from a strong AF interaction as discussed in Sec.~I. In
particular, in Ref.~\cite{Das10} the OC was calculated based on a
self-consistent treatment of the Hubbard model. It was argued
that  the  charge transfer gap observed in the insulating state
of cuprates is due to the AF LRO which splits the CuO$_2$ band
into two magnetic subbands. Assuming that the AF LRO exists at
any doping,  the experimentally observed increase of the charge
transfer energy with doping and a simultaneous decrease of the
MIR absorption energy were explained. In this scenario, the MIR
band  originates from the pseudogap in the electronic spectrum
also induced by the AF LRO. In our theory based on the
consideration of the  $t$--$J$ model we  cannot study the
interesting problem of doping dependence of the charge-transfer
peak at high energy observed in the insulator-to-metal transition
in cuprates. This problem should be considered within the
two-subband Hubbard model as, e.g., in Ref.~\cite{Plakida96},
since in the $t$--$J$ model only the lower Hubbard subband is
explicitly taken into account.
\par
In Ref.~\cite{Vidmar09},  the MIR absorption in the region of
$\omega \approx t - 2t$, being quite strong even without coupling
to phonons, was related  to the interaction of a doped hole with
spin excitations. In our theory, we can explain the MIR absorption
by electron interaction with  spin fluctuations  in the decay of
charge excitations. We cannot relate the MIR absorption to  the
pseudogap in the single-particle electronic spectrum, since the
MIR conductivity shows no notable temperature and doping
dependence which is characteristic for the pseudogap
phenomenon~\cite{Hufner08}.
\par
In the limit of strong correlations the Hubbard model and $t$-$J$
model have been used in calculations of dc and optical
conductivity. As was pointed out in Sec.~I, in numerical studies
of finite clusters, due to a finite energy resolution, only
restricted information on the frequency and temperature
dependence of the OC can be found. The effects of strong
correlations have been efficiently taken into account  within the
DMFT method which enables to  reproduce  qualitatively the main
features of the OC: the Drude peak, the MIR region, the
charge-transfer excitations and the temperature and doping
dependence of the optical spectral weight (see,
e.g.,~\cite{Comanac08,Jarrell95,Toschi05} and references
therein). However, in the DMFT the spatial correlations, such as
short-range AF fluctuations, are not taken into account, and
therefore  the low-energy part of the OC caused by charge-boson
interaction cannot be studied. In the cluster
DMFT~\cite{Haule07}, due to a finite size of clusters, only a
qualitative low-frequency behavior of the OC can be found.  In
the present work the complicated wave-vector dependence of the
dynamical spin susceptibility and the electron interaction with
spin fluctuations are fully taken into account without using
fitting parameters. This  enables to reproduce both the transport
relaxation rate $\Gamma(\omega, T)$ (see Fig.~\ref{figG}) and the
real part of the memory function $M'(\omega, T)$ (see
~Fig.~\ref{figM}), yielding the optical mass renormalization, in
a fair agreement with experiments.
\par
The Allen approximation~\cite{Allen71} for the current-current
correlation function, commonly used in the calculation of  the OC
and the transport relaxation rate $1/\tau$, is based on the
perturbation theory with respect to  $(1/\omega \tau) \ll 1$.  In
our memory-function approach, the optical relaxation rate
(\ref{r3}) describes the direct decay of a charge excitation into
an electron-hole pair assisted by the excitation of spin
fluctuations. In the Allen approach, processes of this type
appear due to  finite life-time effects for the electron-hole
pair. Therefore, to calculate the optical relaxation rate one has
to express the latter in terms of a quasiparticle scattering rate
which is not a straightforward procedure (see, e.g.,
Refs.~\cite{Carbotte05,Schachinger06,Sharapov05}).
\par
The present microscopic theory  has in fact  some limitations
arising from the $t$--$J$ model used in the calculations. Besides
the deficiency of the charge-transfer peak at high energy
discussed above, the complicated structure of the OC found
experimentally in the MIR region  is missed in our theory. This
may be due to polaron effects and the coupling of  magnetic
excitations with phonons via doped holes as discussed in
Refs.~\cite{Cappelluti07,Mishchenko08,Vidmar09}. To overcome these
limitations, an extended Hubbard model including a strong
electron-phonon interaction should be considered.

\section{Conclusion}

In the present paper we have studied the charge dynamics  within
a microscopic theory for  the optical and dc conductivities for
the $t$-$J$ model  by taking into account electron scattering by
spin fluctuations.  In our theory, based on the memory-function
formalism, we calculate directly the transport relaxation rate
without using the Allen perturbation theory.
\par
Within the proposed theory, we are able to obtain a reasonable
agreement with experiments on cuprates for the relaxation rate,
the optical conductivity,  and the  resistivity in broad regions
of temperatures and doping. In particular, in the underdoped
region with a strong AF SRO, a fair quantitative  agreement  was
found for the resistivity, Fig.~\ref{figR_exp} (a), and for the
relaxation rate, Fig.~\ref{figG_YBCO}. This proves the essential
role of AF spin fluctuations in the charge dynamics of cuprates.
This conclusion has been corroborated in a number of theoretical
and experimental studies of OC (see, e.g.,~\cite{Yang09,Heumen09}
and references therein). In the overdoped case, where the AF spin
fluctuations are suppressed, additional scattering mechanisms
(e.g., due to electron-phonon interaction and impurity
scattering) should be invoked to explain experimental data. From
our results we conclude that spin-fluctuations induced by the
kinematic interaction should give a substantial contribution to
the $d$-wave pairing in cuprates as has been shown recently in
Ref.~\cite{Plakida11}.

\acknowledgments

\indent Partial financial support by the Heisenberg--Landau
Program of JINR is acknowledged. One of the authors (N.P.) is
grateful to the  MPIPKS, Dresden, for the hospitality during his
stay at the Institute, where a  part of the present work has been
done.

\appendix

\section{Calculation of the Memory Function}

To derive  Eq.~(\ref{g5}) for the  memory function, we consider
the equations of motion for the relaxation function $\Phi(t-t') =
(\!( J_{x}(t) | J_{x}(t'))\!)$ (see
Refs.~\cite{Goetze72,Plakida96,Plakida97}). Differentiating the
function subsequently over time $t$ and $t'$ we obtain a system
of equations which  in the Fourier representation reads,
\begin{eqnarray}
 \omega \, \Phi(\omega) & = & \chi_{0}+  (\!( F_{x} | J_{x})\!)_{\omega},
 \label{A1a} \\
\omega \,  (\!( F_{x} | J_{x})\!)_{\omega} & = & - (\!( F_{x} |
F_{x} )\!)_{\omega},
 \label{A1b}
\end{eqnarray}
where $\,F_{x} = i\dot{J}_{x} = [ J_{x} , H ]$ is the force
operator. In  Eq.~(\ref{A1b}) the relation of the orthogonality
$\, (F_{x} , J_{x} ) =   (i \dot{J}_{x}, J_{x}) = \langle [ J_{x}
, J_{x} ] \rangle = 0 $ was used.  Introducing the zero-order
relaxation function $\, \Phi_{0}(\omega) =  \chi_{0}/{\omega}$ we
can solve the system of equations (\ref{A1a}) and (\ref{A1b}) in
the form
\begin{equation}
\Phi(\omega)=\Phi_{0}(\omega)- \Phi_{0}(\omega)\, T(\omega) \,
\Phi_{0}(\omega),
 \label{A2}
\end{equation}
with the scattering matrix
\begin{equation}T(\omega) =
(1/\chi_{0}) (\!(F_{x}|F_{x})\!)_{\omega} (1/\chi_{0}) .
 \label{A3}
\end{equation}
The memory function $M(\omega)$ is defined by Eq.~(\ref{g4})
which can be written in the form:
\begin{equation}
\Phi(\omega)=\Phi_{0}(\omega)- \Phi_{0}(\omega)\,
[M(\omega)/\chi_{0}]\, \Phi(\omega).
 \label{A4}
\end{equation}
From Eqs.~(\ref{A2}) and (\ref{A4}) we get a relation between the
memory function and the scattering matrix:
\begin{equation}
T(\omega) = [M(\omega)/\chi_{0}] - [M(\omega)/\chi_{0}]
\Phi_{0}(\omega) T(\omega).
 \label{A5}
\end{equation}
This equation  shows that the memory function is the ``proper
part'' of the scattering matrix (\ref{A3}) which has no parts
connected by a single zero-order relaxation function, $M(\omega)
= \chi_{0} T(\omega)^{\rm proper}$, as given by Eq.~(\ref{g5}).

\section{Mode Coupling Approximation}

Using the spectral representation for the retarded  Green
functions~\cite{Zubarev60}, we write the relaxation rate
(\ref{g8}), $\Gamma(\omega) = M''(\omega) ={\rm
Im}((F_{x}|F_{x}))^{\rm proper}_{\omega +i0^+} (1/\chi_{0})$, in
terms of the time-dependent force-force correlation function:
\begin{equation}
   \Gamma(\omega) = \pi \frac{1- \exp(\beta\omega)}{2\chi_{0}\omega}
   \int_{-\infty}^{\infty} dt e^{i \omega t}
      \langle F_{x} F_{x}(t) \rangle^{\rm proper}.
\label{B1}
\end{equation}
To calculate the force operator $F_{x} =  [ J_{x} , H ]$,  we
first determine the current $\, J_{x} = - i[ P_{x}, H ]$, where
the polarization operator in terms of  HOs reads: $\, P_{x} =
  e \sum_{i} { R}^{x}_{i} \sum_{\sigma}
  X_{i}^{\sigma \sigma}\,$. Using the commutation relations (\ref{r2a})
we derive the expression for the  current operator
\begin{equation}
   J_{x} = i e \sum_{i, j ,\sigma} ( R_{i}^{x} -
R_{j}^{x}) \;   t_{ij} \; X^{\sigma 0}_i X^{0 \sigma}_j .
\label{B2a}
\end{equation}
The force operator describes electron scattering on spin and
charge (density) excitations which  results from the kinematic
interaction for the  HOs. This can be seen from the equation of
motion for the electron annihilation operator:
\begin{eqnarray}
i\frac{d}{dt} X^{0 \sigma}_{i}(t) &= & [X^{0 \sigma}_{i}, H] = -
\mu X^{0 \sigma}_{i} - \sum_{j, \sigma'} \,t_{ij} B_{i \sigma
\sigma'} X^{ 0 \sigma '}_{j}
\nonumber \\
&  + & (1/2)   \sum_{j, \sigma'}\, J_{ij} \; X^{ 0 \sigma '}_{i}
[ B_{j\sigma\sigma'} - \delta_{\sigma'\sigma}]\; ,
  \label{B2}
\end{eqnarray}
where the Bose-like operator $B_{i \sigma \sigma'}$ is introduced,
\begin{eqnarray}
B_{i\sigma\sigma'} & = &
     (X^{00}_{i} + X^{\sigma\sigma}_{i})\delta_{\sigma'\sigma}
  +   X^{\bar{\sigma}\sigma}_{i}\delta_{\sigma' \bar{\sigma}}
\nonumber \\
& = &  [1 - (1/2) N_j + S^z_j ] \delta_{\sigma'\sigma}+
S^{\bar{\sigma}}_j \delta_{\sigma' \bar{\sigma}} . \label{B3}
 \end{eqnarray}
Here  the completeness relation for the HOs and the definition of
the number and spin operators (\ref {r2}) are  used. By this type
of equations of motion, for the force operator we obtain the
expression:
\begin{eqnarray}
&&   F_{x}= -i  e \sum_{i , j , l} \sum_{\sigma\sigma'}
 ( R_{i}^{x} - R_{j}^{x}) \,t_{ij}
\nonumber  \\
&&   \times \; \{\, X^{\sigma 0}_{i} [t_{jl}  X^{ 0 \sigma
'}_{l}B_{j\sigma \sigma'}
 - (1/2)  J_{jl} \; X^{ 0 \sigma '}_{j}  B_{l\sigma\sigma'}]
\nonumber \\
&&- [t_{il}  X^{\sigma'0}_{l} B^\dag_{i \sigma \sigma'}
 - (1/2)  J_{il} \,
  X^{\sigma' 0}_{i}B^\dag_{l\sigma\sigma'}] X^{0 \sigma}_j
\}. \quad \label{B4}
\end{eqnarray}
Introducing the  $\bf q$-representation for HOs and the
interactions,
\begin{eqnarray}
&&X^{0 \sigma}_{i} = \frac{1}{\sqrt{N}} \sum_{{\bf q}}
  X^{0 \sigma}_{ \bf q} \;  e^{ i{\bf q R}_{i}},
\nonumber \\
&&  B_{j\sigma\sigma'} = \frac{1}{N} \sum_{{\bf q}}
   B_{{\bf q} \sigma\sigma'} \; e^{ i{\bf q R}_{j}},  \quad
\label{B4a} \\
&& t_{ij} = \frac{1}{N}\sum_{{\bf q}} t({ \bf q}) \,
   e^{ i{\bf q R}_{ij}}, \;
 J_{ij} =\frac{1}{N}\sum_{{\bf q}}  J({ \bf q}) \,
   e^{ i{\bf q R}_{ij}},
\nonumber
\end{eqnarray}
where ${\bf  R}_{ij} = {\bf R}_{i} - {\bf R}_{j}$,  the force
operator (\ref{B4}) takes the form,
\begin{eqnarray}
&& F_{x} = - \frac {e}{N} \sum_{\bf k , q}
   \sum_{\sigma\sigma'} \, v_{x}({\bf k}) \; [t({\bf k-q})
 -(1/2) J({\bf q})]\,
 \nonumber \\
&& \times \{X^{\sigma 0}_{\bf k} X^{ 0 \sigma '}_{\bf k-q}
 \,B_{{\bf q} \sigma \sigma^{'}} -
 X^{\sigma' 0}_{\bf  k -q } X^{ 0 \sigma }_{\bf k}
 \,B_{-{\bf q} \sigma^{'} \sigma}\},
 \label{B5}
\end{eqnarray}
 where $\, v_{x}({\bf k}) = - \partial {t({\bf k})}/ \partial
{k_{x}} $ is the electron velocity. Changing the indexes in the
last term,
 ${\bf k'=  k -q }, \, \sigma
\leftrightarrow \sigma'$, and ${\bf  q}\rightarrow - {\bf q}$,
 we obtain the final expression
\begin{eqnarray}
&& F_{x}  =  - \frac{e}{N} \sum_{\bf k, q} \sum_{\sigma\sigma'}\{
\,v_x({\bf k})\,
 [t({\bf k-q}) - (1/2)J({\bf q}) ]
\nonumber  \\
&& -  \,v_x({\bf k-q})\,
 [t({\bf k}) - (1/2)J({\bf q}) ]\}
 X^{\sigma 0}_{\bf  k} X^{0 \sigma' }_{\bf k - q} B_{{\bf
q}\sigma\sigma'}\nonumber \\
 &&\equiv - \frac {e}{N} \sum_{\bf k , q}
   \sum_{\sigma\sigma'}\,g_{x}({\bf k,k-q})\,
 X^{\sigma 0}_{\bf k} X^{ 0 \sigma '}_{\bf k-q}
 \,B_{{\bf q} \sigma \sigma^{'}}. \label{B5a}
\end{eqnarray}
In the last equation we introduce the transport vertex $g_{x}({\bf
k,k-q})$ given by Eq.~(\ref{r4}).
\par
We calculate the many-particle time-dependent correlation
functions in Eq.~(\ref{B1}) in the mode-coupling approximation
assuming  an independent propagation of electron and charge-spin
excitations. In this approximation, the time-dependent correlation
functions can be written as a product of fermionic and bosonic
correlation functions:
\begin{eqnarray}
 &&\langle X^{{\sigma}0}_{\bf k} X^{0 \sigma'}_{\bf k-q}
  B_{{\bf q} \sigma \sigma'} |
 X^{\sigma'0}_{\bf k-q}(t) X^{0\sigma}_{\bf k} (t)
B^{\dagger}_{{\bf q} \sigma \sigma'}(t)\rangle
\label{B6}\\
& = &
 \langle X^{{\sigma}0}_{\bf k} X^{0{\sigma}}_{\bf k} (t) \rangle
 \langle X^{0 \sigma'}_{\bf k-q} X^{\sigma' 0}_{\bf k-q}(t)\rangle
 \langle B_{{\bf q} \sigma\sigma'}
 B^{\dagger}_{{\bf q} \sigma\sigma'}(t) \rangle \; .
\nonumber
\end{eqnarray}
Using  the definition for  the Bose-like operator  (\ref{B3}), for
the bosonic correlation function we obtain
\begin{eqnarray}
&& \langle B_{{\bf q} \sigma\sigma'}
 B^{\dagger}_{{\bf q} \sigma\sigma'}(t) \rangle =
\langle \{ [X^{00}_{{\bf q}} + X^{\sigma\sigma}_{{\bf q}}]\,
\delta_{\sigma'\sigma}
  +   X^{\bar{\sigma}\sigma}_{{\bf q}}\,\delta_{\sigma' \bar{\sigma}}\} \,
\nonumber \\
&& \times \{[X^{00}_{-{\bf q}}(t) + X^{\sigma\sigma}_{-{\bf
q}}(t)]\, \delta_{\sigma'\sigma}
  +   X^{\sigma \bar{\sigma}}_{-{\bf q}}(t)\,\delta_{\sigma' \bar{\sigma}}\}\rangle
 \nonumber \\
&&= \langle \{ X^{00}_{{\bf q}} + X^{\sigma\sigma}_{{\bf q}}\}\,
\{ X^{00}_{-{\bf q}}(t) + X^{\sigma\sigma}_{-{\bf q}}(t)\}
\rangle \, \delta_{\sigma'\sigma}
\nonumber \\
&& +  \langle  X^{\bar{\sigma}\sigma}_{{\bf q}} X^{\sigma
\bar{\sigma}}_{-{\bf q}}(t) \rangle\, \delta_{\sigma'
\bar{\sigma}}
   =  (1/4)\langle N_{{\bf q}}N_{-{\bf q}}(t) \rangle \, \delta_{\sigma'\sigma}
 \nonumber \\
&&  +
 \langle  S^z_{{\bf q}}| S^z_{-{\bf q}}(t)
\rangle \, \delta_{\sigma'\sigma}
 + \langle X^{\bar{\sigma}\sigma}_{{\bf q}}X^{\sigma \bar{\sigma}}_{-{\bf q}}(t)
\rangle\, \delta_{\sigma' \bar{\sigma}}.
 \label{B7}
\end{eqnarray}
In the paramagnetic state, for the sum of the spin correlation
functions in Eq.~(\ref{B7}) we have: $\langle S^z_{{\bf q}}
S^z_{-{\bf q}}(t) \rangle + \langle S^{-}_{{\bf q}}S^{+}_{-{\bf
q}}(t) \rangle = \langle {\bf S}_{{\bf q}}{\bf S}_{-{\bf q}}(t)
\rangle $.  Finally, using spectral representations for the
time-dependent correlation functions in
Eq.~(\ref{B6}),~\cite{Zubarev60}
\begin{equation}
\langle B A(t)\rangle =\int_{-\infty}^{\infty} d\omega
\,e^{-i\omega t} f(\omega)[-(1/\pi)]{\rm Im} \langle \!\langle
A|B \rangle\!\rangle_{\omega}, \nonumber
\end{equation}
where $f(\omega)$ is the Fermi function $n(\omega)$ for the
correlation function $ \langle X^{{\sigma}0}_{\bf k}
X^{0{\sigma}}_{\bf k} (t) \rangle$ and  the Bose function
$N(\omega)$ for the charge-spin correlation functions, after
integration over time $t$ in Eq.~(\ref{B1}) we obtain the
expression (\ref{r3}) for the relaxation rate.

\end{document}